\documentclass [aps,prb,10pt,twocolumn,showpacs,superscriptaddress]{revtex4-1}
\usepackage[english]{babel}
\usepackage{graphicx}
\usepackage{amsmath,amssymb}
\usepackage[utf8]{inputenc}
\usepackage{natbib}
\usepackage{units}
\usepackage{hyperref}
\usepackage{color,soul}
\usepackage{url}
\hypersetup{colorlinks=true,citecolor={blue},linkcolor={blue},urlcolor={blue}}
\usepackage{balance}

\newif\iffig\figfalse

\begin{document}
\title{Polariton Spin Whirls}

\date{\today}

\author{P. Cilibrizzi}
\affiliation{School of Physics and Astronomy, University of Southampton, Southampton, SO17
 1BJ, United Kingdom}

\author{H. Sigurdsson}
\affiliation{Division of Physics and Applied Physics, Nanyang Technological University 637371, Singapore}
\affiliation{Science Institute, University of Iceland, Dunhagi-3, IS-107 Reykjavik, Iceland}

\author{T.C.H. Liew}
\affiliation{Division of Physics and Applied Physics, Nanyang Technological University 637371, Singapore}

\author{H. Ohadi}
\affiliation{School of Physics and Astronomy, University of Southampton, Southampton, SO17
 1BJ, United Kingdom}

\author{S. Wilkinson}
\affiliation{School of Physics and Astronomy, University of Southampton, Southampton, SO17
 1BJ, United Kingdom}

\author{A. Askitopoulos}
\affiliation{School of Physics and Astronomy, University of Southampton, Southampton, SO17
 1BJ, United Kingdom}

\author{I. A. Shelykh}
\affiliation{Division of Physics and Applied Physics, Nanyang Technological University 637371, Singapore}
\affiliation{Science Institute, University of Iceland, Dunhagi-3, IS-107 Reykjavik, Iceland}
\affiliation{ITMO University, St. Petersburg, 197101, Russia}

%\author{P.G. Savvidis}
%\affiliation{FORTH-IESL, P.O. Box 1385, 71110 Heraklion, Crete, Greece}
%\affiliation{Department of Materials Science and Technology, University of Crete, 71003 Heraklion, Crete, Greece}

\author{P. G. Lagoudakis}
%\email[correspondence address:~]{pavlos.lagoudakis@soton.ac.uk}
\affiliation{School of Physics and Astronomy, University of Southampton, Southampton, SO17
 1BJ, United Kingdom}

\begin{abstract}
We report on the observation of spin whirls in a radially expanding polariton condensate formed under non-resonant optical excitation. Real space imaging of polarization- and time-resolved photoluminescence reveal a spiralling polarization pattern in the plane of the microcavity. Simulations of the spatiotemporal dynamics of a spinor condensate reveal the crucial role of polariton interactions with a spinor exciton reservoir. Harnessing spin dependent interactions between the exciton reservoir and polariton condensates allows for the manipulation of spin currents and the realization of dynamic collective spin effects in solid state systems. 
\end{abstract}

\pacs{}
\maketitle

\section{Introduction}
\label{sec:1}

Phase transitions in atomic Bose Einstein condensates (BECs) are associated with symmetry breaking and the appearance of topological defects. In quantum fluids the appearance of quantized vortices in a rotating condensate marks the transition to the superfluid regime\,\cite{butts_predicted_1999}. In the case of spinor condensates the extra degree of freedom provided by the spin, gives rise to complex spin patterns, known as merons\,\cite{Wilson_2013} and skyrmions\,\cite{Leslie_2009}. These structures appear as intricate spin textures due to the rotation of the spins across the condensate induced by dipole-dipole interactions\,\cite{Wilson_2013}. Spontaneous rotation of the spin textures and breaking of chiral symmetry has been reported in a spinor BEC with ferromagnetic interactions\,\cite{saito_2006}. Skyrmions and other nontrivial spin structures have also been observed in 2D superfluid Fermi gas \cite{dong_fermi_superfluid_2015}, topological insulators \cite{hsieh_topological_insulators_2009} and magnetic thin film materials \cite{yu_thin_films_2010}. This tremendous interest in exploring the physics of spin textures is motivated by their strong relation with fundamental phenomena, such as the spin Hall effect in semiconductors \cite{dyakonov_spin_Hall_1971,kato_observation_2004} and spontaneous symmetry breaking in BECs \cite{sadler_spontaneous_BEC_2006}, but also by their potential in future applications, such as low-power magnetic data storage \cite{shibata_towards_2013} and logic devices \cite{fert_skyrmions_2013}.  

In this work, a dynamical spin texture in polariton microcavity is studied for the first time. Polaritons are bosonic quasiparticles formed by the strong coupling between quantum well excitons and the photonic mode of a planar semiconductor microcavity\,\cite{kavokin_microcavities_2007, byrnes_exciton-polariton_2014}. When the polariton population is increased above a threshold density, stimulated scattering leads polaritons to macroscopically occupy the ground state of the dispersion and form a \textit{nonequilibrium} BEC \cite{kasprzak_bose-einstein_2006,Wouters_2008}. % characterized by an inversion-less amplification of the polariton emission \cite{Imamoglu_polariton_laser_1996}. %The resulting macroscopic ground state population is a \textit{nonequilibrium} BEC \cite{kasprzak_bose-einstein_2006}. %and its steady-state results from a dynamical balance of pumping and losses \cite{Wouters_2008}. 
%Compared to conventional condensed matter systems, polariton condensates in semiconductor microcavities provide a unique opportunity to study and characterize spinor dynamics since their spin can be optically accessed by means of polarization measurements. Polaritons possess a spin with two possible projections of the angular momentum $(S_z=\pm1)$ on the structural growth axis $(z)$ of the microcavity. %Superpositions of $S_z=\pm1$ states give rise to linear or elliptical polarization of polaritons\,\cite{Glazov_2013}. 
Being bosons, polaritons possess an integer spin with two possible projections of the angular momentum $(S_z=\pm1)$ on the structural growth axis $(z)$ of the microcavity. Their spin can be optically accessed by means of polarization measurements and described theoretically within the pseudospin formalism \cite{Kavokin_Lagoudakis_2003}. One of the most important effects involving polariton spin, is the so called optical spin Hall effect (OSHE)\,\cite{kavokin_optical_2005}, observed in both polaritonic\,\cite{leyder_observation_2007} and photonic\,\cite{maragkou_optical_2011} microcavities.\,The effect is enabled by the energy splitting between transverse-electric (TE) and transverse-magnetic (TM) polarized modes\,\cite{panzarini_exciton_1999}, which occurs naturally in microcavities and results in spin currents propagating over hundreds of microns in both resonant\,\cite{langbein_polarization_2007} and non-resonant configurations\,\cite{kammann_nonlinear_2012}. Due to the OSHE, the long range coherence\,\cite{long_range_2013} and fast spin dynamics\,\cite{lagoudakis_stimulated_2002}, polaritons have been proposed as a potential candidate for the realization of a new generation of spinoptronic devices\,\cite{liew_polaritonic_2011}. In this regard, the contribution of a spin dependent exciton reservoir has not been considered thoroughly, although in non-resonant experiments and in the proximity of the excitation spot, exciton interactions dominate over other types of interactions \cite{DeGiorgi_relaxation_oscillation_2014} and can directly affect the spin dynamics of polaritons \cite{gao_spin_2015}.
%In our work, we show for the first time that this interaction can result in the peculiar rotation of the spin textures (i.e. the spin whirl) on a picoseconds scale. 
%An important step toward polariton based spin devices was the prediction\,\cite{kavokin_optical_2005} and observation of the optical spin Hall effect (OSHE), in both polaritonic\,\cite{leyder_observation_2007} and photonic\,\cite{maragkou_optical_2011} microcavities.\,The effect is enabled by the energy splitting between transverse-electric (TE) and transverse-magnetic (TM) polarized modes\,\cite{panzarini_exciton_1999}, which occurs naturally in microcavities and results in spin currents propagating over hundreds of microns in both resonant\,\cite{langbein_polarization_2007} and non-resonant configurations\,\cite{kammann_nonlinear_2012}. %In the latter case, the polarization of the pump played a crucial role in the formation of spin patterns since it determined the spin population of the condensate\,\cite{Hamid_2012}.
%
%
\begin{figure*}[!hbtp]
  \centering
  	\includegraphics[scale=0.42]{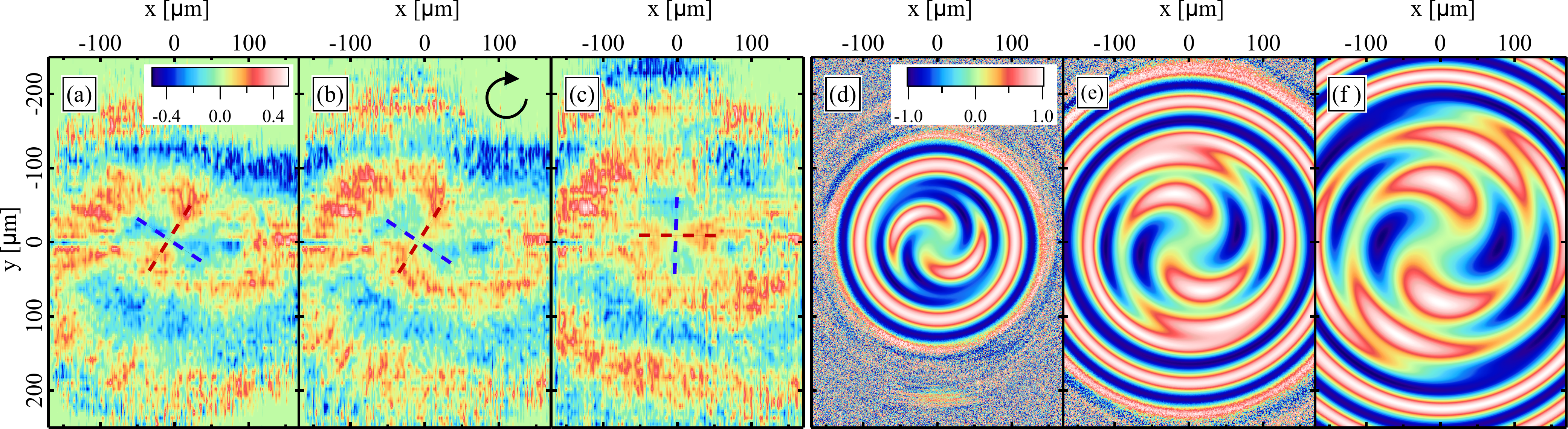}
\caption{Snapshots of the spatio-temporal dynamics of the degree of circular polarization $S_z$ under non-resonant linearly polarised excitation at: (a) $\unit[38]{ps}$, (b) $\unit[41]{ps}$ and (c) $\unit[46]{ps}$ showing the clockwise rotation of the spin texture within the microcavity plane (zero time is defined at the PL onset, see the full dynamics in supplementary video S1). (d-f) Theoretical simulations showing the circular Stokes vector of the spin whirls at: (d) $\unit[30]{ps}$, (e) $\unit[45]{ps}$ and (f) $\unit[60]{ps}$. The parameters used in the simulations are reported in Ref. [34].}
\label{fig:1}
\end{figure*}

In this Letter, we report on the experimental observation of spin whirls in the radial expansion of a polariton condensate formed under non-resonant optical excitation in a GaAs quantum well (QW) microcavity. A spin whirl is a spin texture that rotates in the microcavity plane due to the interplay between the TE-TM splitting and the interaction with an exciton reservoir. The TE-TM splitting alone is responsible for the formation of symmetric 2D spin textures, which is intrinsically a linear effect\,\cite{langbein_polarization_2007}. As a consequence, the orientation of the spin current in the microcavity plane remains fixed in time. However, in the case of a radially expanding condensate, nonlinear interactions with the exciton reservoir at the spatial center of the condensate produce a spiralling effect, which culminates in a coherent rotation of the whole spin texture. The rotation is traced in an inherent spin imbalance in the exciton reservoir that acts as an effective magnetic field due to the anisotropic exciton-polariton interactions. We observe the spiralling effect in both time- and energy-resolved measurements. Simulations based on the Gross-Pitaevskii equation (GPE) coupled with an exciton reservoir unveil the role of the spin imbalanced exciton reservoir in reproducing the experimental observations. 

The paper is organized as follows: In Sec.~\ref{sec:2} we describe the sample and the experimental setup. In Sec.~\ref{sec:3} we report the main experimental and theoretical results showing the rotating spin textures in the plane of the microcavity, i.e. the spin whirls. In Sec.~\ref{sec:4} the theoretical model is presented and explained. In Sec.~\ref{sec:5} we discuss the physical origin of the spin whirls and present additional measurements. Conclusions and perspectives are reported in Sec.~\ref{sec:6}.

%%%%%%%%%%%%%%%%%%%%%%%%%%%%%%%%%%%%%%%%%%%%%%
\section{Sample and experimental setup}
\label{sec:2}

We use a $5 \lambda/2$ AlGaAs/GaAs microcavity, with four sets of three QWs, characterized by a cavity photon lifetime of ${\sim}\,\unit[9]{ps}$ and a Rabi splitting of $\unit[9]{meV}$. All the data presented here are recorded at negative exciton-photon detuning, $\Delta{=}\unit[-4]{meV}$, and under a non-resonant ($\unit[1.653]{eV}$) pulsed optical excitation ($\unit[250]{fs}$, $\unit[80]{MHz}$) of $\unit[7]{m W}$, focused to a $\sim\unit[2]{~\mu m}$ FWHM spot using a 0.4 numerical aperture objective. The excitation beam is linearly polarised with polarization extinction ratio higher than $1{:}10^3$. Photoluminescence (PL) is then collected in reflection geometry through the same objective, analyzed by a polarimeter composed of a $\lambda/2$ or $\lambda/4$ waveplate and a linear polarizer and projected on the entrance slit of a streak camera, with $\unit[2]{ps}$ temporal resolution (see section 1 (S1) of the supplementary information\,\cite{suppl_info} for details). 

%%%%%%%%%%%%%%%%%%%%%%%%%%%%%%%%%%%%%%%%%%%%%
\section[Rotating Spin Texture: The Spin Whirls]{Rotating Spin Texture: \\The Spin Whirls}
\label{sec:3}

%\subsection{Spin Whirls}
%\label{subsec:3a}

Under non-resonant linearly polarized excitation, time and polarization resolved measurements reveal a clockwise rotation of the entire spin texture in the plane of the microcavity at an angular velocity of about $\unit[0.11]{rad/ps}$. This is shown in Figs.\,\ref{fig:1} (a-c) for the $z$-component of the Stokes vector, i.e., the degree of circular polarization, $S_z$=$(I_{\psi_+}-I_{\psi_-})/I_{tot}$, with $I_{\psi_+}$ and $I_{\psi_-}$ being the measured intensity of the two circular polarization components and $I_{tot}=I_{\psi_+}+I_{\psi_-}$ (see supplementary information\,\cite{suppl_info}, video S1 for full dynamics).\,The non-resonant excitation creates a reservoir of hot excitons, which rapidly relaxes to populate the lower polariton dispersion and form a polariton condensate\,\cite{kasprzak_bose-einstein_2006}.\,At the pump spot position, due to the repulsive interactions between polaritons and the exciton reservoir, the condensate is blueshifted in energy. Outside the pump spot, this potential energy is converted to kinetic energy with an in-plane wavevector (here k $\leq\unit[2.8] {\mu m^{-1}}$) determined by the cavity lifetime and the gradient of the potential\,\cite{Wouters_2008}. Thus, highly focused Gaussian excitation ($\sim\unit[2]{~\mu m}$ FWHM), produces a cylindrically symmetric potential that leads to the radial expansion of polaritons in the plane of the microcavity (see video S2). 

\section{Theoretical model}
\label{sec:4}

To accurately model the spin dynamics in the exciton-polariton system, an open-dissipative Gross-Pitaevskii equation (1) describes the polariton spinor order parameter ($\Psi_\pm$), which is then coupled with the exciton reservoir density ($\mathcal{N}_\pm$) \,\cite{PhysRevLett.99.140402}: 
\begin{align} \notag
i \hbar \frac{d \Psi_{\pm}}{d t} &= \Big[ \hat{E}  - \frac{i \hbar }{2 \tau_p} + \alpha |\Psi_{\pm}|^2 + G \sigma_\pm P(\mathbf{r},t) \\
& + \Big( g_R + \frac{i \hbar r_c}{2} \Big) \mathcal{N}_{\pm} \Big] \Psi_{\pm} + \hat{H}_{\text{LT}} \Psi_{\mp} ,
\end{align}
\begin{equation}
\frac{d \mathcal{N}_{\pm}}{d t} = - \left( \frac{1}{\tau_x} + r_c |\Psi_{\pm}|^2 \right) \mathcal{N}_{\pm} + \sigma_{\pm} P(\mathbf{r},t).
\end{equation}
These equations model the process of polaritons being generated from a hot exciton reservoir and then scattered into the ground state of the condensate. The coupled equations take into account the energy blueshift of the condensate due to interactions with excitons (with interaction strength $g_R$). $\hat{E}$ is the condensate kinetic energy, $\tau_p$ and $\tau_x$ are the polariton and exciton lifetimes respectively. It has been shown that the dominant component of interactions between polaritons comes from the exchange interaction \cite{Ciuti_exchange_1998}. In our model, the same-spin polariton interactions strength is characterized by the parameter $\alpha$. We neglect interactions between polaritons with opposite spins, which are typically smaller in magnitude\,\cite{Vladimirova_2010} at energies far from the biexciton resonance\,\cite{Takemura_2014}. The exciton reservoir is driven by a Gaussian pump, $P(\mathbf{r},t)$, as in the experiment, and feeds the polariton condensate with a condensation rate $(r_c)$. An additional pump-induced shift is described by the interaction constant $G$ to take into account other excitonic contribution to the blueshift\,\cite{PhysRevLett.99.140402}. The polarization of the pump is controlled by the parameters $\sigma_+$ and $\sigma_-$ (e.g., a horizontally polarized pump would correspond to $\sigma_+ = \sigma_- = 1$). $\hat{H}_{\text{LT}}$ is the TE-TM splitting which mixes the spins of the polaritons:
\begin{equation}
\hat{H}_{\text{LT}} = \frac{\Delta_{LT}}{k_{LT}^2}  \left( i \frac{\partial}{\partial x} \pm  \frac{\partial}{\partial y} \right)^2,
\end{equation}
with $\Delta_{LT}$ being half TE-TM splitting at wavevector $k_{LT}$. The TE-TM splitting is defined by the ratio $\Delta_{LT}/k_{LT}^2$, while the in-plane wavevector of polaritons is given by the operator in the round brackets. The parameters used in the simulations are reported in Ref. \cite{parameters}.
 %In all calculations the following parameters were set to: $\alpha = 2.4$ $\mu$eV $\mu$m$^2$, $g_R = 1.5 \alpha$, $G = 4 \alpha$, $r_c = 0.01$ $\mu$m$^2$ ps$^{-1}$, $\Delta_{LT}/k_{LT}^2 = 11.9$ $\mu$eV $\mu$m$^2$, $\tau_p = 9$ ps, $\tau_x = 10$ ps.

\section{Discussion}
\label{sec:5}

\subsection{Different pump polarizations give rise to different spin textures}
\label{subsec:5a}

\begin{figure}[!hbtp]
  \centering
		\includegraphics[scale=0.38]{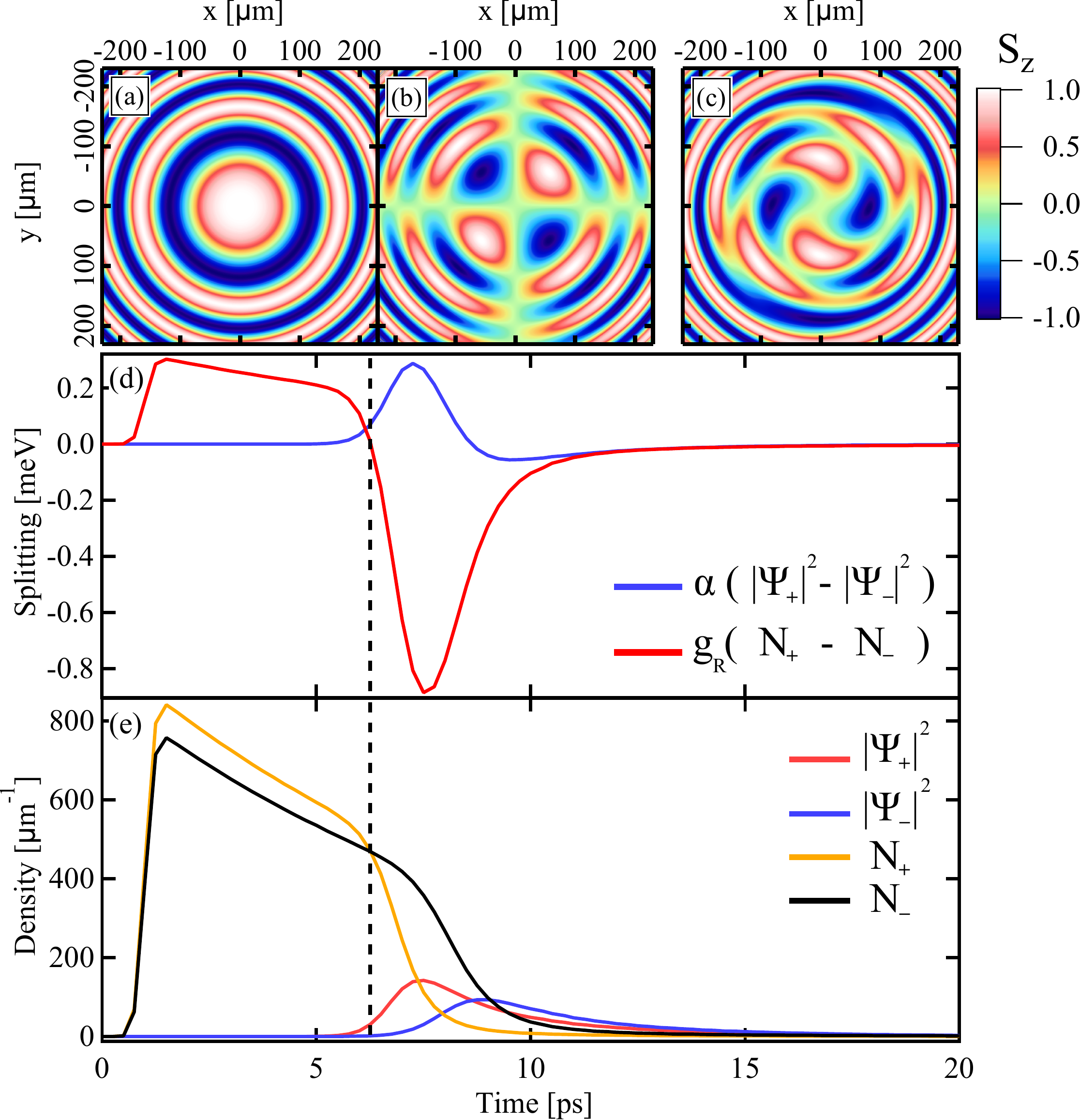}
	\caption{(a-c) Spin textures showing the evolution of the degree of circular polarization $S_z$ after $\unit[50]{ps}$ in a system excited with (a) nearly circular ($\sigma_+{=}\,1; \sigma_-{=}\,0.1$), (b) linear ($\sigma_+{=}\,\sigma_-{=}\,1$) and (c) elliptical ($\sigma_+{=}\,1; \sigma_-{=}\,0.9$) pump polarization. (d-e) Dynamics of the condensate and reservoir at the pump position under elliptical pumping. (d) Energy splitting versus time of the polariton condensate (blue line) and exciton reservoir (red line). (e) Density versus time of the $\Psi_{\pm}$ polariton condensate and $\mathcal{N}_{\pm}$ exciton reservoir at the pump position. The dashed line indicates the position where the energy splitting in (d) reverse.}
\label{fig.th.pulse}
\end{figure}

In Figure \ref{fig.th.pulse}, the theoretical circular Stokes polarization patterns obtained with circularly [Fig.\,\ref{fig.th.pulse}\,(a)] and linearly [Fig.\,\ref{fig.th.pulse}\,(b)] polarized pump are shown. %and elliptically (Fig.\,\ref{fig.th.pulse} c) 
In the nonlinear regime, the circular pump allows for injection of a single spin condensate that due to OSHE evolves to concentric rings of alternating spin\,\cite{kammann_nonlinear_2012}, as shown in Fig.\,\ref{fig.th.pulse}\,(a). The cylindrically symmetric patterns observed here are due to the fact that the polariton pseudospin, directed along the $z$-axis in the Poincar\'e sphere, is always perpendicular to the effective magnetic field, lying on the $x$-$y$ plane. Under linearly polarized pump [Fig.\,\ref{fig.th.pulse}\,(b)] due to the absence of spin imbalance in the exciton reservoir, the fermionic component of excitons produces strong exchange coupling between bright and dark states that force the condensate to be linearly polarized \cite{Combescot_2007}. In this case, the typical OSHE pattern is retrieved due to the Stokes vector precessing at $45^{\circ}$ to the $x$, $y$ axis\,\cite{kavokin_optical_2005}. It has been predicted that under linear excitation the condensate forms a Skyrmion pattern\,\cite{flayac_transmutation_2013}. %This prediction was preceded by the experimental observation of similar polarization patterns\,\cite{langbein_polarization_2007}.

For the creation of polarisation symmetry breaking textures such as the spin whirls observed here, a spin imbalance is necessary. Although we excite with a highly linearly polarized beam (extinction ratio higher than $1:10^3$), an ellipticity is created due to the high numerical aperture of the focusing lens. Indeed, the electric field of a linearly polarized beam, when focused by a high-NA objective, acquires non-zero components in the two directions perpendicular to the polarization of the incident field (i.e., at the focal plane the electric field vector sweeps an ellipse)\,\cite{richards_electromagnetic_1959,chen_chapter_2012}. Thus, the tight focus of a linearly polarized excitation beam, breaks the rotational symmetry of the $\sigma_+$ and $\sigma_-$ polarizations and introduces an ellipticity in the pump spot. We have measured an ellipticity of $10\%$ for the excitation conditions used in the experiment (see supplementary information\,\cite{suppl_info}, S3). 
\newline
In the simulations, we introduce a $10\%$ ellipticity in the linearly polarized pump, i.e., elliptical pulse with $(\sigma_+,\sigma_-){=}(1,0.9)$, and observe that the circular polarization patterns rotate, as shown in Fig.\,\ref{fig.th.pulse}\,(c). 

\subsection{Spin whirls origin}
\label{subsec:5b}

To understand this behavior, we must first consider that polaritons can only be generated in the vicinity of the localized pump spot, which serves as the source for the entire spatial spin pattern. The time-dependent spatial rotation observed in our configuration is, in fact, a manifestation of varying polarization at the pump spot location. %Experimentally, this is confirmed by the fact that calculate the degree of circular polarization at the pump spot position oscillates between the average values of $\pm 0.1$, as shown in Fig.\,\ref{fig:3}\,(a).  
\begin{figure}[!hbtp]
  \centering
		\includegraphics[scale=0.4]{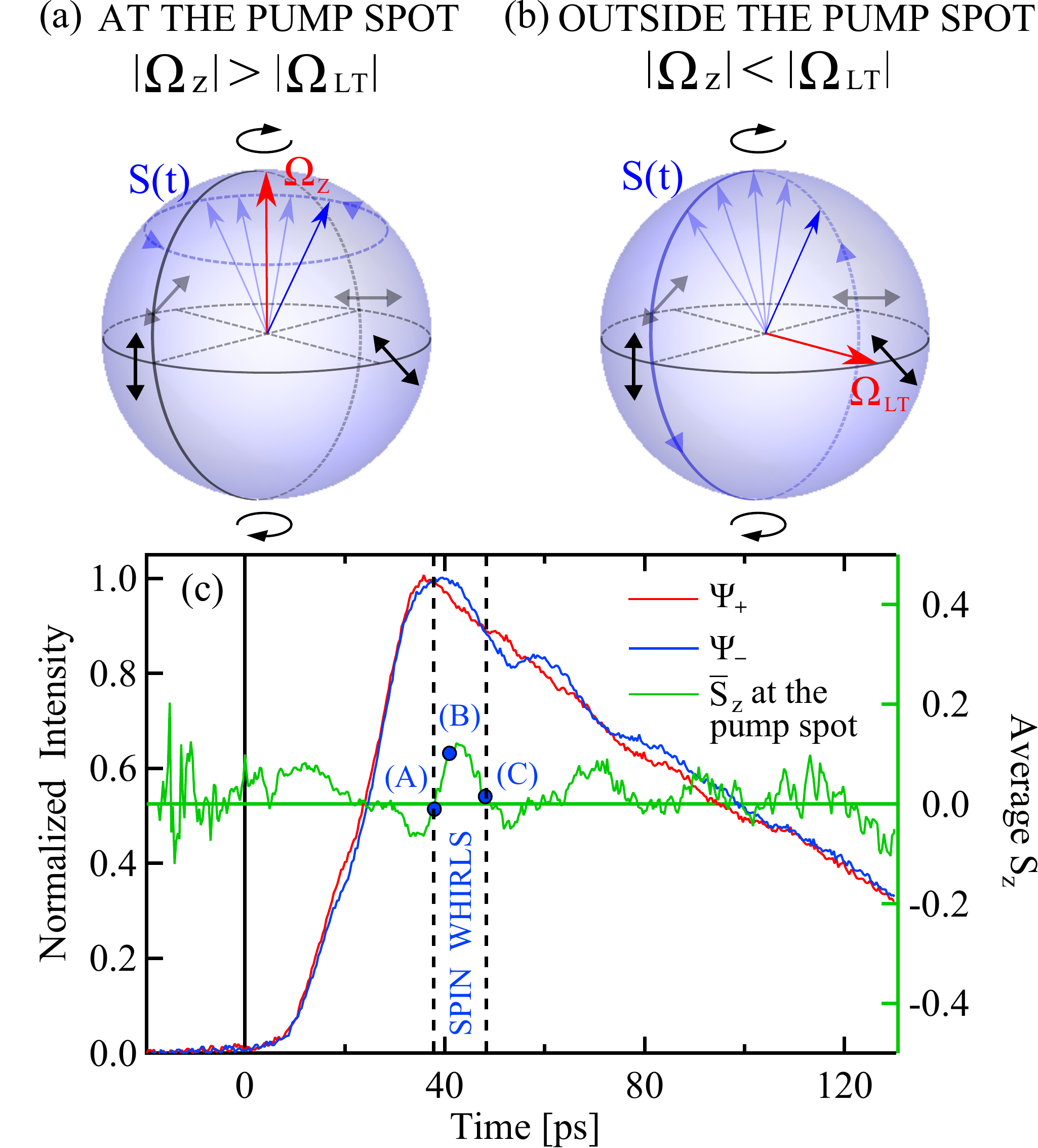}
		\caption{The pseudospin vector S(t) (blue arrows) in the Poincar\'e sphere at: (a) the pump spot and (b) outside the pump. At the pump spot position, (a), S(t) precesses around the $z$-direction since $\left|\Omega_z\right|>\left|\Omega_{LT}\right|$. Outside the pump spot, (b), S(t) precess around $\Omega_{LT}$ since $\left|\Omega_{LT}\right|>\left|\Omega_z\right|$.(c) Time-resolved, spatially integrated measurements of the two circular polarization components ($\Psi_+$, red and $\Psi_-$, blue) PL intensity, normalized and integrated over the area imaged in Figs.\,\ref{fig:1}(a-c), i.e., $(\unit[460]x\,\unit[340]){\mu m}^2$.\,In green we show the time resolved degree of circular polarization $S_z$ averaged over an area ($\unit[1.78] x \unit[1.78]){\mu m}^2$, centered at $(\unit[0,0]){\mu m}$ in Figs.\,\ref{fig:1}(a-c), comparable with the $2~\mu m$ FWHM excitation spot.\,The blue solid circles annotated with (A), (B), (C) refer to the three snapshots of Figs.\,\ref{fig:1}(a-c).}
\label{fig:3}
\end{figure}

The varying polarization at the pump spot is generated by the ellipticity of the Gaussian pump, which populates one circular component of the reservoir faster than the other. This leads to a splitting $g_R(\mathcal{N}_+ - \mathcal{N}_-)$ of polaritons (Fig.\,\ref{fig.th.pulse}(d)), which can be thought of as an effective Zeeman splitting at the pump spot. Here, the imbalance between the two populations [Fig.\,\ref{fig.th.pulse}\,(e)] induces an effective magnetic field along the $z$-direction ($\Omega_z$)\,\cite{Renucci_2005}, which causes the precession of the Stokes vector in the Poincar\'e sphere, as shown schematically in Fig.\,\ref{fig:3}\,(a). Due to its excitonic nature, $\Omega_z$ exists only at the pump spot position where the exciton reservoir is localized. Away from the excitation spot, the polariton pseudospin dynamics is essentially driven by the TE-TM splitting of the polariton mode, represented by an in-plane effective magnetic field, $\Omega_{LT}$\,\cite{kavokin_optical_2005} [Fig.\,\ref{fig:3}(b)]. The combination of these two rotations is at the origin of the polariton spin whirls. The rotating polarization at the source results in the appearance of rotating spiral arms in the spatial distribution of the circular polarization degree, in analogy to the water jets created by a rotating sprinkler head [Figs.\,\ref{fig:1}\,(d-f)]. The energy splitting between $\Psi_+$ and $\Psi_-$ states at the pump spot can also be generated by interactions between polaritons, $\alpha (|\Psi_+|^2 - |\Psi_-|^2)$, where the corresponding precession in linear polarization was previously described\,\cite{shelykh_semiconductor_2004}, however, we find that the dominant contribution to the splitting is caused by the exciton reservoir splitting, $g_R(\mathcal{N}_+ - \mathcal{N}_-)$ (see supplementary information\,\cite{suppl_info}, S4). The small imbalance between $\Psi_+$ and $\Psi_-$, induced by the ellipticity of the pump polarization, results in picosecond scale oscillation in the circular emission [red and blue profile in Fig.\,\ref{fig:3}\,(c)] indicated in the literature as features of bosonic stimulation \cite{Martin_2002,shelykh_spin_2005}. Experimentally, the rotation of the polarization at the pump spot is confirmed by the average of the degree of circular polarization calculated at the pump spot position, which oscillates between $\pm 0.1$, as shown in Fig.\,\ref{fig:3}\,(c) (green profile) and coincides with the rotation of the spin textures [Figs.\ref{fig:1}\,(a-c)]. The differences in the time dynamics observed in experiment and theory are due to differences in the reservoir dynamics occurring at the pump spot position. Typically polariton condensation is described with the use of a single reservoir model\,\cite{Wouters_2008}. While modeling using multiple reservoir levels may offer a closer fit to the dynamics\,\cite{Lagoudakis_2011,Carlos_E_relaxation_2013}, we do not expect significant changes in the spatial patterns, which are the main focus of our work.

%\subsection{Circularly polarized excitation}
%\label{subsec:5c}
\subsection{Additional measurements}
\label{subsec:5c}

We have repeated the same experiments at the same conditions of detuning, power and excitation spot-size but now exciting with a circularly polarized beam (see supplementary information\,\cite{suppl_info},\,S5 and video S3). In this case, polariton condensation results in highly imbalanced population [Fig.\,S4\,(b)] and the small ellipticity induced by the tightly focused spot will not play a relevant role as in the case of linearly polarized pump. As a consequence, the imbalance between the two polariton populations is set by the pump and preserved throughout the entire process so that no oscillation of the polarization appear at the pump spot [Fig.\,S6\,(d)] and the spin texture does not rotate (see supplementary information\,\cite{suppl_info}, S5). 
\begin{figure}[!hbtp]
\includegraphics[scale=0.45]{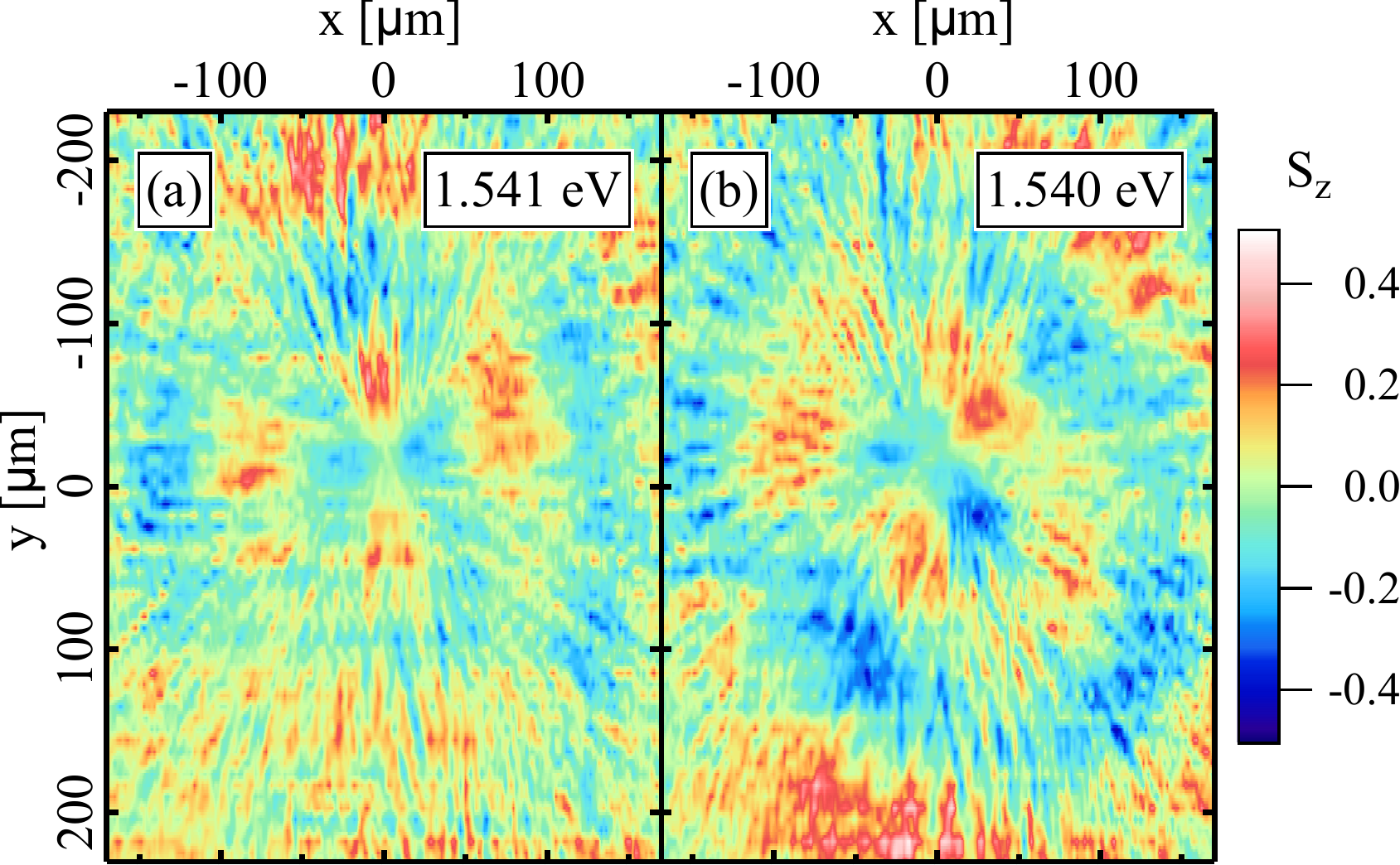}
\caption{Snapshots of real space, spectral tomography of the degree of circular polarization $S_z$ at: (a) $\unit[1.541]{eV}$ and (b) $\unit[1.540]{eV}$ showing the clockwise rotation of the spin whirls within the microcavity plane under non-resonant linearly polarized excitation.} 
\label{fig:4}
\end{figure}
%
%

%\subsection{Energy resolved measurements}
%\label{subsec:5d}

Finally, we study the rotation of the spin textures using real-space spectral tomography under the same excitation conditions as in Fig.\,\ref{fig:1}\,(a-c).\,This is shown in Fig.\,\ref{fig:4}. Under non-resonant optical excitation, the pseudospin dynamics of polaritons is strongly connected with the energy relaxation of the exciton reservoir immediately after the arrival of the excitation pulse. The decay of the exciton reservoir in time results to a gradual decreasing potential energy that polaritons experience at the pump spot. %Thus, spin textures observed at later times correspond to those detected at lower energies. 
The interplay between the polariton spin and the energy relaxation of the exciton reservoir give rise to spin textures with different chirality (i.e., their image does not coincide with their respective mirror image) at different energies, similarly to the spin vortices at different energy observed in atomic condensate with ferromagnetic interactions\,\cite{saito_2006}. The typical quadrature of the OSHE rotates by ${\sim}\,45^{\circ}$ in the plane of the microcavity due to the rotation of the linear polarization axis by ${\sim}\,90^{\circ}$ in the Poincar\'e sphere. This is also confirmed in k-space (see supplementary information\,\cite{suppl_info}, video S4), where the variation of the linear polarization at the source results in the appearance of rings of opposite circular polarization\,\cite{amo_anisotropic_2009}. Thus, due to the varying polarization at the pump spot and the decrease of the blue shift with time, polaritons with spin up/down populate concentric rings in k-space (see supplementary information\,\cite{suppl_info}, video S4). 

\section{Conclusions}
\label{sec:6}

In conclusion,\,we have observed and studied the dynamics of spin whirls in polariton microcavities. We demonstrated that the appearance of spin whirls is due to a dynamical optical spin Hall effect, which originates from the TE-TM splitting of propagating modes and a self-induced Zeeman splitting at the pump spot. The strong nonlinear interactions between polaritons and the exciton reservoir induce a collective rotation of the 2D textures in the plane of the microcavity. An analogous but static pattern of indirect exciton spin currents was observed under continuous wave excitation and a real magnetic field in coupled quantum wells\,\cite{high_spincurrents_2013}. Here, we emphasize the dynamic induction of an effective magnetic field on the picosecond scale and the resulting dynamic control of spin currents, which is an additional step toward the realization of spinoptronic devices.
\\
\section*{ACKNOWLEDGMENTS}
\label{sec:7}

P.C. and P.G.L. acknowledge P.G. Savvidis and Z. Hatzopoulos for providing the sample. P.C. and P.G.L. acknowledge support by the Engineering and Physical Sciences Research Council, UK [Project EP/M025330/1]. The data from this paper can be obtained from the University of Southampton e-Print research repository at \url{http://dx.doi.org/10.5258/SOTON/384780}. H.S. and I.S. acknowledge support by the FP7 IRSES POLATER, ITN NOTEDEV and Rannis ``Bose, Fermi and hybrid systems for spintronics''.
%%%%%%%%%%%%%%%%%%%%%%%%%%%%%%%%%%%%%%%%%%%%%%
%%%%%%%%%%%%%%%%%%%%%%%%%%%%%%%%%%%%%%%%%%%%%%
%%%%%%%% SUPPLEMENTARY INFORMATION %%%%%%%%%%%
%%%%%%%%%%%%%%%%%%%%%%%%%%%%%%%%%%%%%%%%%%%%%%
%%%%%%%%%%%%%%%%%%%%%%%%%%%%%%%%%%%%%%%%%%%%%%
\setcounter{figure}{0} \renewcommand{\thefigure}{\textbf{S\arabic{figure}}}
\renewcommand{\thesection}{S\arabic{section}}
\setcounter{section}{0}

%\cleardoublepage
\begin{widetext}

\begin{center}
{\large\textbf{Supplementary Information}}
\end{center}

\hspace{1.0cm}

\section{Sample \& Experimental Setup}
\label{sec:1}

The sample used is a $5 \lambda/2$ AlGaAs/GaAs microcavity, composed by 32 (35) top (bottom) distributed Bragg reflectors (DBRs) and 4 triplets of $\unit[10]{nm}$ thick GaAs QWs. The cavity quality factor is measured to exceed $Q \gtrsim 8000$, with transfer matrix simulations giving $Q = 20000$, corresponding to a cavity photon lifetime $\sim\unit[9]{ps}$. The Rabi splitting is $\unit[9]{meV}$. This is the the same sample used in Ref.~\cite{kammann_nonlinear_2012}. All the data presented here are recorded at negative detuning $\Delta =\unit[-4]{meV}$. 
\begin{figure}[!hbtp]
\centering
\includegraphics[scale=0.6]{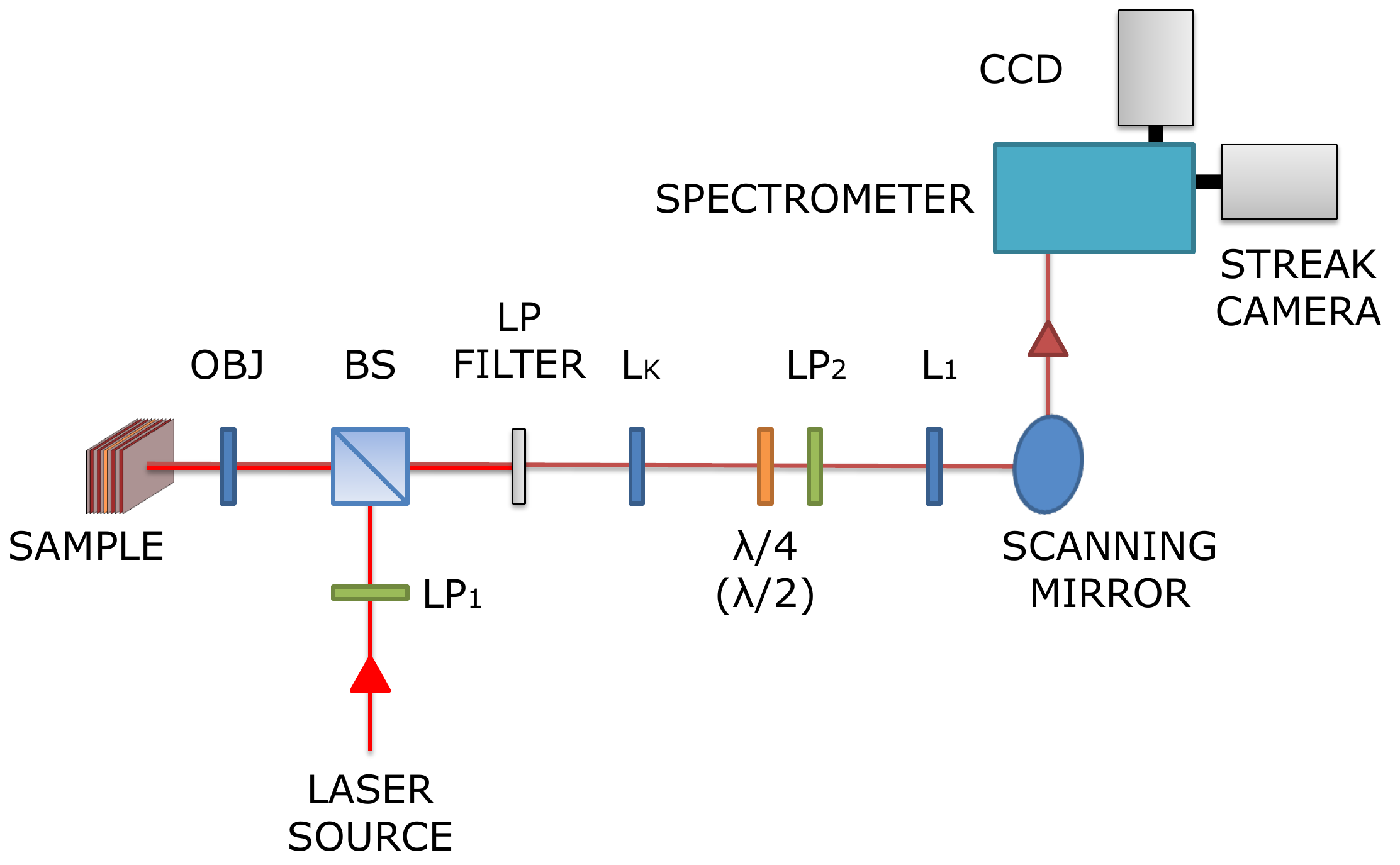}
\caption{ Sketch of the experimental setup. Lists of the optical components: \textbf{LP$_1$} is the linear polarizer with extinction ratio higher than $1000{:}1$; \textbf{BS} is the non-polarizing beam splitter; \textbf{OBJ} is the 20x, 0.4 NA objective; \textbf{LP Filter} is the long pass filter; \textbf{L$_k$} is the lens used to imagine the k-space; \textbf{$\lambda/4$} (\textbf{$\lambda/2$}) is the quarter-wave plate (half-wave plate); \textbf{LP$_2$} is a linear polarizer and $\textbf{L1}$ is the $\unit[10]{cm}$ focal length lens. The motorized scanning mirror and the spectrometer, equipped with both a charge coupled device (CCD) and a streak camera, are also shown.  }
\label{fig:1}
\end{figure}
\\
The dynamic of polaritons is studied by using the experimental setup shown in Fig. \ref{fig:1}. We use a pulsed laser, with pulse width of $\unit[250]{fs}$ and a pulse frequency of $\unit[80]{MHz}$. The sample is held in a helium cryostat at temperature $T= 6$ K and all the experiments are performed under non-resonant excitation. The excitation laser, tuned to the first spectral minimum above the high-reflectivity mirror stopband at $\unit[1.653]{eV}$, is focused to a spot of $\sim\unit[2]{~\mu m}$ FWHM $(2\sqrt{\ln{(2)}}\sigma)$ (with $\sigma$ being the standard deviation of the 1D Gaussian curve used to fit the spot intensity profile), by means of a 0.4 numerical aperture objective. The power of the excitation is at $\unit[7]{m W}$. The excitation beam is linearly polarised with polarization extinction ratio higher than $1{:}10^3$.The polarized emission is collected by the same objective, sent through a long pass filter (to filter out the laser), analyzed by a polarimeter composed of a $\lambda/2$ or $\lambda/4$ plate and a linear polarizer (LP$_2$) and then projected with an achromatic lens L$_1$ (f = $\unit[10]{cm}$) on the entrance slit of a spectrometer. The latter is equipped with both a charge coupled device (CCD) and a streak camera with $\unit[2]{ps}$ temporal resolution. The grating of the spectrometer can be interchanged with a mirror allowing direct imaging of real or momentum (k) space without energy resolution. To measure the far field emission of the microcavity, i.e., the k-space, an extra lens L$_k$ (f = $\unit[30]{cm}$) is used to form the image of the Fourier plane onto the entrance slit of the spectrometer. 
\\
\\
In order to study the polarization dynamics in both real and k-space, without energy resolution, the intensity emitted by the microcavity is time-resolved by using a tomography scanning technique. In this technique, the near field intensity I(t,x,y) [far field I(t,k$_x$,k$_y$)] emitted by the microcavity is imaged on the entrance slit of the streak camera for a fixed $y$ [k$_y$]. By using a motorized mirror, it is possible to scan the y [k$_y$] dimension and acquire I(t,x) [I(t,k$_x$)] at different y [k$_y$]. In this way, a 2D real space image I(x,y) [2D k-space image I(k$_x$,k$_y$)] at different times can be reconstructed. 
In the case of k-space this is possible since photons, emitted from the microcavity at an angle $\theta$ (where $\theta$ is the emission angle with respect to the normal of the microcavity plane), correspond to polaritons with in-plane wavevector k$_{||}$ = k$_0$sin($\theta$) (with k$_0=2\pi/\lambda_0$ and $\lambda_0$ being the emission wavelength). The data reconstruct with this technique are Fig.1 (a-c) of the manuscript, Fig.\ref{fig:C_stokes} (a-c) and the videos (S1-S3) for the 2D real space and the video S4 for the k-space.
In order to take into account the different delay introduced by the waveplates, when the different polarization components are measured, the experimental data have been rescaled to have a common zero, where zero time is defined at the photoluminescence onset. This correction has been applied to both the Stokes images and the intensity profiles presented in this work. 

\noindent
Finally, the tomography scanning technique has been also used to perform energy resolved measurements of the spin texture [Fig.4 (a-b) of the manuscript]. In this case the emission from the microcavity is focused on the entrance slit of the spectrometer with the grating now active (for the spectral selection of the ultrashort pulses) and then projected on the CCD. The motorized mirror allow to scan the wavelength ($\lambda$) on the grating and acquire I(x,y) at different $\lambda$. Successively, a 2D real space image I(x,y) at different $\lambda$ can be reconstructed. 

%We use the grating of the monochromator to obtain an image of the polariton disperssion ($\theta$, E), where $\theta$ is the emission angle woith respect to the normal of the microcavity plane. On the other hand, to measure the k-ring we use the monochromator as a mirror to image the Fourier plane on the entrance slit of the Streak camera. In this way a k-ring image ($\theta, \phi$), with $\phi$ being the azimuth angle can be obtained.

\section{Spin whirls in presence of disorders}
\label{sec:disorder}

In Fig.\ref{fig:1b}(a-c), the formation of polariton spin whirls is calculated in presence of disorders, resembling the experimental results shown in Figs.1(a-c) of the main manuscript. The parameters used to perform the simulations are the same used for Figs.1(a-c) of the main manuscript \cite{parameters}. The disorder
potential was generated with 0.05 meV root mean squared amplitude and 1.5 $\mu m$ correlation length. The theoretical calculations show that, although disorder introduces some additional fine structure, it does not affect the basic spin textures.
\begin{figure}[!hbtp]
\includegraphics[scale=0.6]{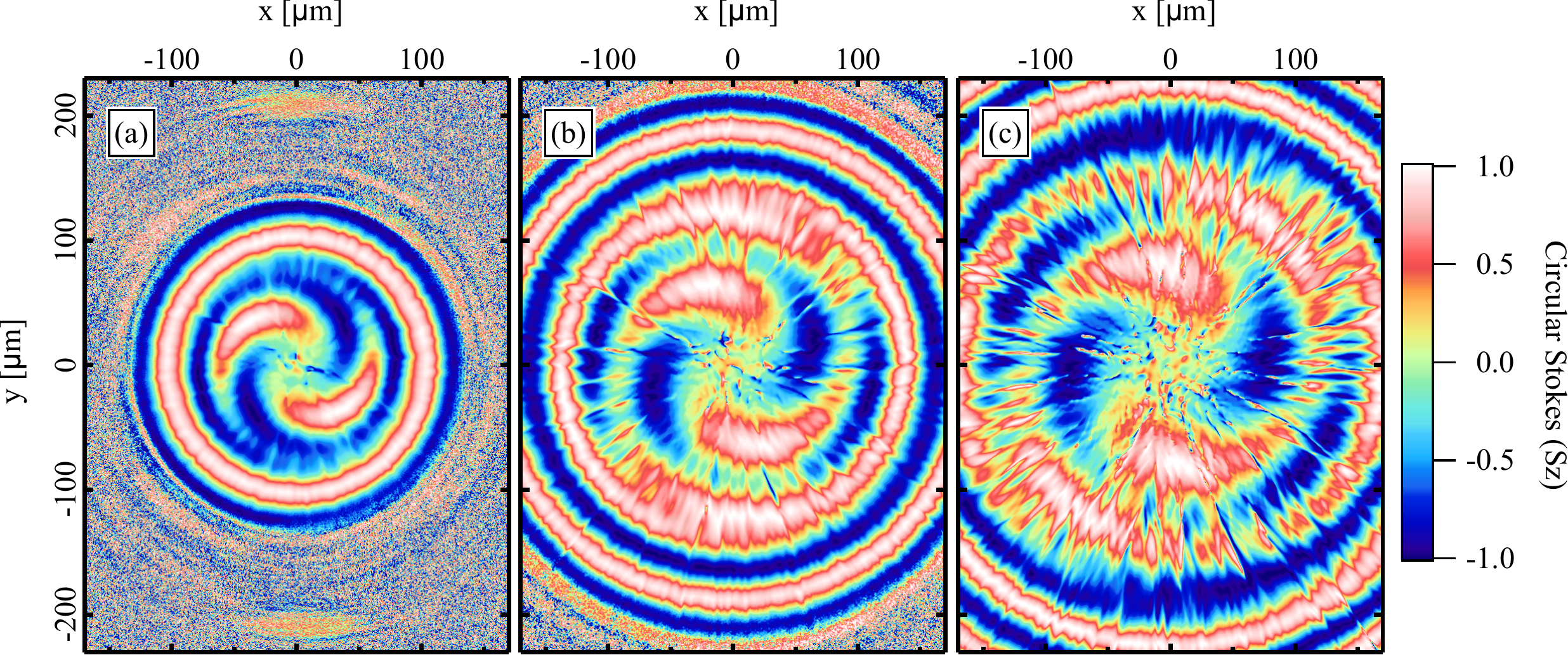}
\caption{(a-c) Theoretical simulations showing the circular Stokes vector $S_z$ of the spin whirls in presence of disorders at (a) $\unit[30]{ps}$, (b) $\unit[45]{ps}$ and (c) $\unit[60]{ps}$. The parameters used in the simulations, reported in Ref.\cite{parameters}, are the same as Fig.1(d-f) of the main manuscript.}
\label{fig:1b}
\end{figure}
%
%

%\newpage
\section{Ellipticity of the excitation spot}
\label{sec:2}
As has been shown by Richards and Wolf \cite{richards_electromagnetic_1959}, the tight focusing of a polarized Gaussian beam through a high numerical aperture (NA) lens, results in the modification of the polarization at focal plane. In particular, the electric field of a linearly polarized beam, when focused by a high NA objective, acquires non-zero components in the two directions perpendicular to the polarization of the incident field (i.e., at the focus plane the electric field vector sweeps an ellipse) \cite{richards_electromagnetic_1959,lekner_polarization_2003,dorn_focus_2003,kang_polarization_2010}. For a review see Ref. \cite{chen_chapter_2012}.
In our experiments, the polarization at the pump spot has been measured by focusing a linearly polarized beam with a 0.4 NA objective (the same used in the experiment) on a glass. The transmitted intensity has been collected with a 100x, 0.7 NA objective. In order to measure the polarization of the beam, we used a polarimeter composed of a $\lambda/2$ or $\lambda/4$ plate and a linear polarizer. The emission is then imaged in real space by a 20 cm focus lens directly on a CCD camera. The linear (S$_x$), diagonal (S$_y$) and circular (S$_z$) Stokes parameters measured are reported respectively in Figs. \ref{fig:2} (a-c). 
\begin{figure}[!hbtp]
\includegraphics[scale=0.6]{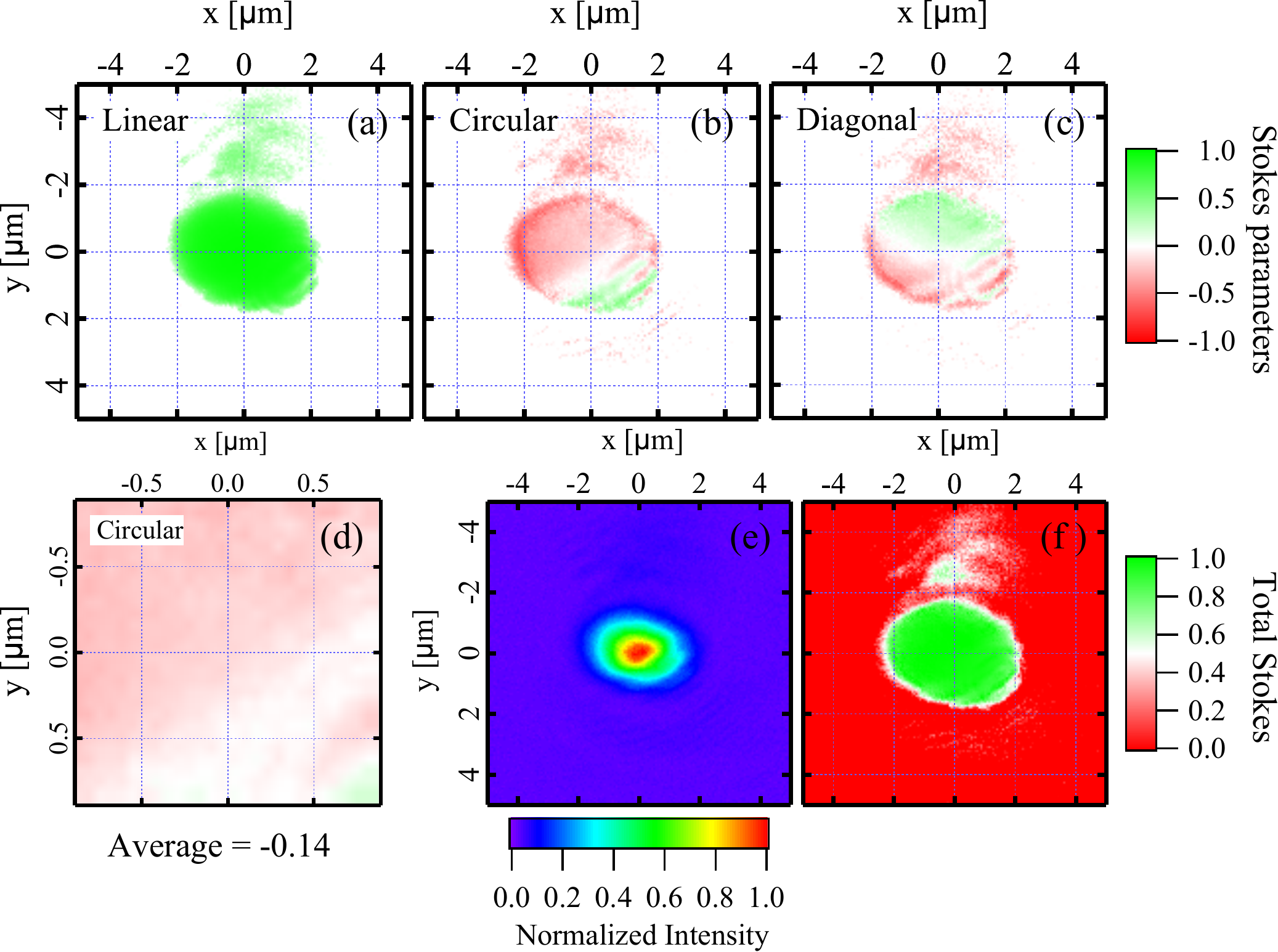}
\caption{(a) Linear, (b) circular and (c) diagonal Stokes parameters of the beam spot focused by a 0.4 numerical aperture objective. (d) Zoom of the circular component reported in (b) on an area of $(\unit[1.78] x \unit[1.78]){\mu m}^2$  comparable with the area of the $2~\mu m$ FWHM excitation spot used in the experiment. The average of the circular Stoke components is $-0.14$. (e) Real space intensity of the circularly polarized pump spot ($\sigma_-$). (f) Total degree of polarization $S_{tot}$ calculated from (a), (b) and (c).}
\label{fig:2}
\end{figure}

\noindent
The Stoke parameters S$_{x,y,z} = (I_{H,D,\Psi_+}-I_{V,A,\Psi_-})/I_{tot}$, with I$_{H,D,\Psi_+}$ and I$_{V,A,\Psi_-}$ being the measured intensity in the horizontal (H), vertical (V), diagonal (D), antidiagonal (A) and the two circular ($\Psi_+ , \Psi_-$) polarization components. $I_{tot}=I_{H,D,\Psi_+}+I_{V,A,\Psi_-}$ is the total emission.
We estimate the circular value of ellipticity by averaging the circular Stokes parameters over an area of $(\unit[1.78] x \unit[1.78]){\mu m}^2$ comparable with the area of the $2~\mu m$ FWHM excitation spot used in the experiment, as shown in Fig.\ref{fig:2} d). The average value of the circular stokes parameters is $-0.14$ justifying the $10\%$ ellipticity used in the theoretical simulations. %Similar values of ellipticity have been reported in Refs (find the references). 
In Fig.\ref{fig:2} e) and Fig.\ref{fig:2} f) the intensity of the beam in real space for the circular component ($\sigma_-$) and the total degree of polarization calculated as S$_{tot} = \sqrt{S_x^2+S_y^2+S_z^2}$ are also reported for the sake of completeness.

\section{Polariton and Exciton reservoir Dynamics}
\label{sec:3a}

In Fig.\ref{fig:5}, the intensity of the polarization components integrated over space is plotted versus time in the case of linearly polarized pump and circular detection [Fig.\ref{fig:5} (a)] and circularly polarized pump and linearly detection [Fig.\ref{fig:5} (b)], corresponding respectively to the intensity profiles of Fig.1 of the main manuscript and Fig.\ref{fig:C_stokes} of the supplementary information (see Sec.\ref{sec:3}).
In both cases the area of integration is $(\unit[460] x~\unit[340]){\mu m}^2$. To take into account the different delays introduced by the waveplates, the intensity profiles have been scaled in order to have a common zero. In the case of linearly polarized pump [Fig.\ref{fig:5} (a)] the small imbalance between $\Psi_+$ and $\Psi_-$ introduced by the high-NA objective, results in picosecond scale oscillation in the circularly polarized emission (similarly to the one observed in Ref. \cite{Martin_2002}), which coincides with the rotation of the spin textures. %In general the effective magnetic field acting on the polariton pseudospin has two components, one directed along the z axes and the other lying in the plane of the microcavity. %The beats between the circularly polarized components are due to the in-plane component of the effective magnetic field \cite{deveaud_book_2007}.
\begin{figure}[!hbtp]
\includegraphics[scale=0.54]{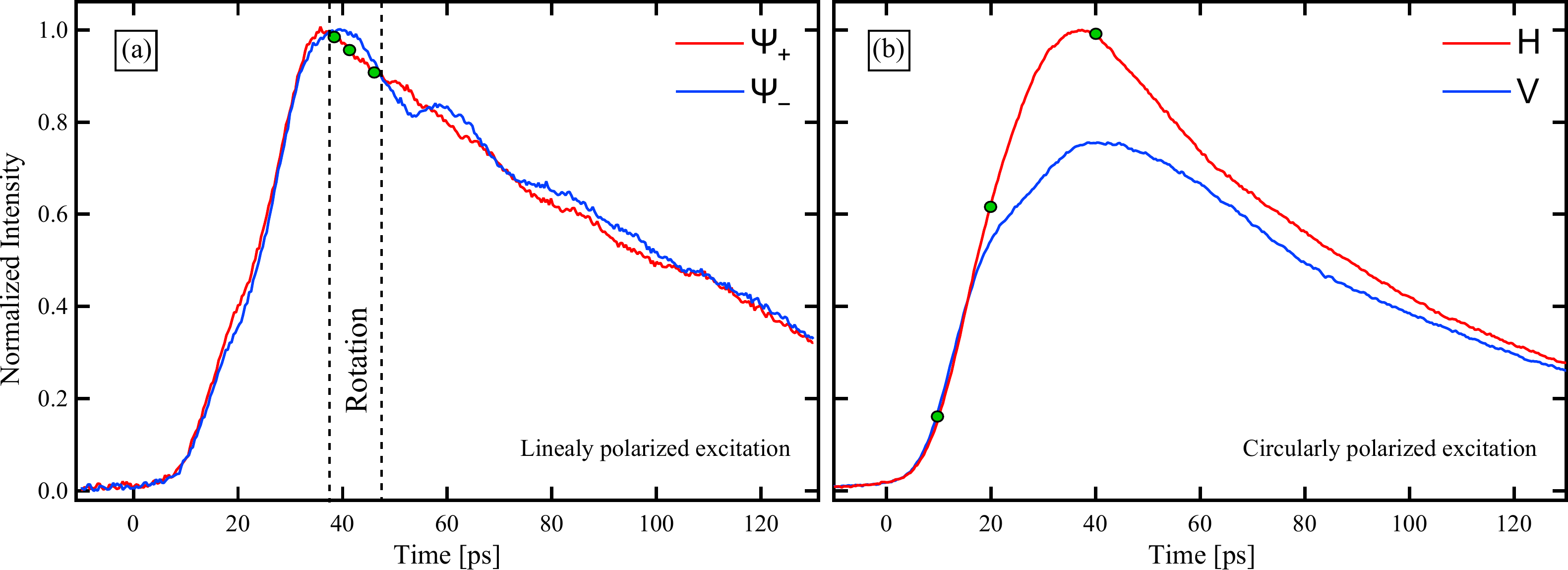}
\caption{Time-resolved, spatially integrated measurements under (a) linearly polarized excitation and circular detection and (b) circularly polarized excitation and linear detection. In both cases the intensity profiles versus time have been integrated over an area $(\unit[460] x \unit[340]){\mu m}^2$ corresponding to the one imaged in Fig.1 of the manuscript and Fig.\ref{fig:C_stokes} of the supplementary information. The solid green circles in (a) and (b) correspond respectively to the time where the snapshots in Fig.1 (a-c) of the manuscript and Fig.\ref{fig:C_stokes} (a-c) of the supplementary information have been extracted.}
\label{fig:5}
\end{figure}

\noindent
Therefore, by referring to Fig. \ref{fig:5} (a), we can identify three different regimes in the formation dynamics of the spin whirl. Up to $\unit[38]{ps}$, the dynamics of the polariton is mainly characterized by the propagation of polaritons radially out of the excitation spot (see the supplementary video S2) which corresponds to the formation of a 2D spin texture in the plane of the microcavity. Once the 2D spin textures are formed, corresponding to the point of maximum intensity of $\Psi_+$ and $\Psi_-$ polaritons, they start to rotate. This regime, indicated in Fig. \ref{fig:5} (a) as "Rotation regime", corresponds to the appearance of the spin whirl [Figs. 1 (a-c) of the main manuscript] and the inversion of the polarization at the pump spot position [Fig. 3(a) of the main manuscript]. In the case of circularly polarized excitation, on the other hand, the imbalance between the two polariton populations set by the pump will be preserved throughout the entire process and no oscillations in the density appear [Fig. \ref{fig:5} (b)]. 
\\
\\
In Fig.\ref{fig.cent} the dynamics of the polariton condensate and exciton reservoir [Fig.\ref{fig.cent} (a)], under elliptically polarized pump, is studied and compared with the energy splitting [Fig.\ref{fig.cent} (b)] and the polarization at the pump spot position [Fig.\ref{fig.cent} (c)]. Starting from a linearly polarized condensate shown in Fig.\ref{fig.cent} (a), the evolution of the polarization follows hand in hand with the splitting [Fig.\ref{fig.cent} (b)]. Specifically, one should note the sudden change in the polarization behavior as the splitting reverses (grey dashed line), corresponding to the Stokes vector reversing its precession in Fig.\ref{fig.cent} (c). The rotational direction is controlled by the sign of the splitting. Positive splitting induces anti-clockwise rotation while negative splitting a clockwise one. When $\Psi_+$ polaritons are generated faster they deplete the $\mathcal{N}_+$ excitons, causing the density to suddenly drop faster than $\mathcal{N}_-$ and thus changing the sign of the splitting. Numerically, the spin $+1$ polaritons condense first since they are being pumped at a higher rate due to the ellipticity of the pump. This is displayed as a concentric polarization ring which expands outward. Then, as $S_z = -1$ polaritons condense and the reservoir densities deplete (i.e. the splitting switches from weak positive to strong negative) the spin whirl appears.%Qualitatively the rotation observed in experiment is the same as depicted by the GPE model. To achieve more accurate dynamics, a multiple-reservoir model would be more suitable.
\begin{figure}[!hbtp]
  \centering
		\includegraphics[scale=0.8]{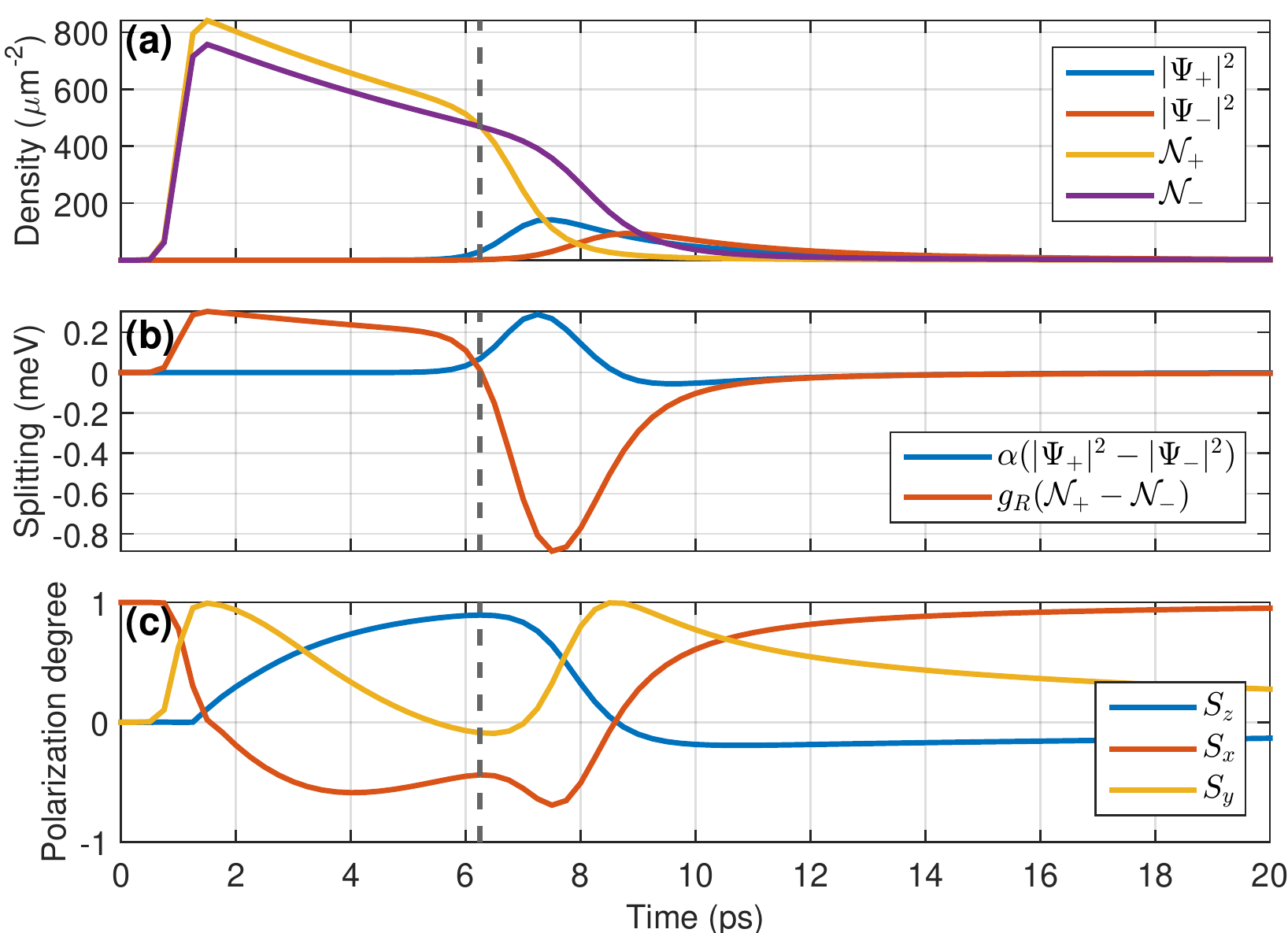}
	\caption{Dynamics of the condensate and reservoir at the pump position under elliptical pumping extracted from the simulation in Fig.2 (c) of the main manuscript. (a) Density versus time of the $\Psi_{\pm}$ polariton condensate and $\mathcal{N}_{\pm}$ exciton reservoir. (b) Energy splitting versus time of the polariton condensate (blue line) and exciton reservoir (orange line). (c) Linear ($S_x$), diagonal ($S_y$) and circular ($S_z$) Stokes components versus time, extracted at the pump spot position. The dashed grey line indicates the position where the energy splitting in (b) and the Stoke vector in (c) reverse.}
\label{fig.cent}
\end{figure}

A notable difference exists between the experiment and the simulation. Experimentally, as noted earlier, there is an equal and steady formation of the spin components (under elliptical pumping) which at their maximum intensity suddenly rotate [Fig. \ref{fig:5} (a)]. From the GPE model [Eqs. (1) and (2), main manuscript] the imbalance set by the pump immediately causes an imbalance in the exciton reservoir components, $\mathcal{N}_\pm$. When the polaritons condense, the $\Psi_\pm$ populations grow at different rates [blue and red line in Fig.\ref{fig.cent} (a) unlike the steady growth seen in Fig.\ref{fig:5} (a)] leading up to the whirl. This causes an immediate appearance of the spin whirl in the simulations from the moment of condensation and splitting reversal (Fig.\ref{fig.cent}, gray dashed line). Qualitatively the rotation observed in experiment is the same as depicted by the GPE model. To achieve more accurate dynamics, a multiple-reservoir model would be more suitable. 
\\
\\
\textbf{Single Reservoir Model.} Typically polariton condensation is described with the use of a single reservoir model \cite{Wouters_2008}. This approach is known to result in an exaggerated depletion of the reservoir, which is emptied once condensation is stimulated. Accurate descriptions of polariton condensate dynamics require the multi-level structure of the reservoir to be accounted for \cite{Lagoudakis_2011,Carlos_E_relaxation_2013}. However, the single reservoir model is able to predict the spatial pattern of the spin whirl, which is the main focus of our work, and its qualitative rotation in time. To avoid using an overly complicated model to describe this effect, we prefer the single reservoir model, while sacrificing an exact match to the timescales observed experimentally (in Fig.\ref{fig.cent}).  %We notice that the rotation of the 2D textures coincides with the onset of oscillations in intensity between the two circular polarization components. Afterward, the intensity of polariton decays and any remarkable effect disappear. 

%The rotational direction is controlled by the sign of the splitting. Positive splitting induces anti-clockwise rotation while negative splitting a clockwise one. When $\Psi_+$ polaritons are generated faster they deplete the $\mathcal{N}_+$ excitons, causing the density to drop suddenly faster than $\mathcal{N}_-$ and thus changing the sign of the splitting (gray vertical dashed line in Fig. S3 b). Numerically, the spin $+1$ polaritons condense first since they are being pumped at a higher rate due to the ellipticity of the pump. This is displayed as a concentric polarization ring which expands outward. Then, as $S_z = -1$ polaritons condense and the reservoir densities deplete (i.e. the splitting switches from weak positive to strong negative) the spin whirl appears.

%\newpage
\section{Circularly polarized excitation}
\label{sec:3}

As mentioned in the main manuscript, we have repeated the same experiments at the same conditions of detuning ($-\unit[4]{meV}$), power ($\unit[7]{m W}$) and excitation spot ($\unit[2]{\mu m}$ FWHM) but now exciting with a circularly polarized beam. The experimental results are shown in Fig.\ref{fig:C_stokes} and in the supplementary video S3.
\begin{figure}[!hbtp]
\centering
\vspace{0.1cm}
\includegraphics[scale=0.5]{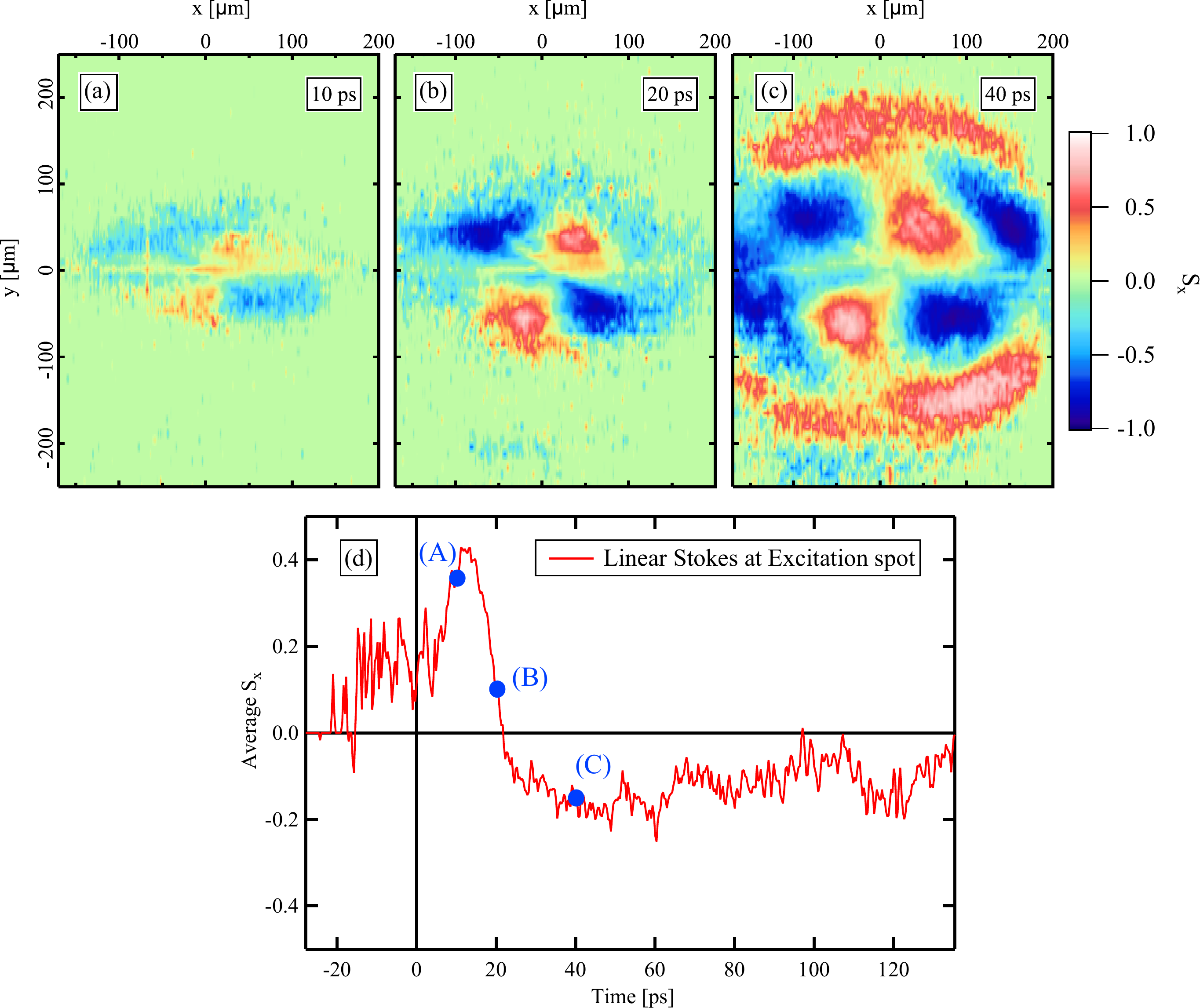}
\caption{Real space experimental Stokes parameters at (a) $\unit[10]{ps}$, (b) $\unit[20]{ps}$ and (c) $\unit[40]{ps}$ showing the evolution in time of the linear  component within the microcavity plane. The excitation beam is circularly polarized and at $\unit[1.653]{eV}$ energy. After the hot excitons relax down on the lower polariton dispersion, polaritons are formed with k $\leq\unit[2.8] {\mu m^{-1}}$. See supplementary video S3 for the full dynamics in real space. (d) Linear degree of polarization versus time calculated by averaging the experimental circular Stokes parameters over an area of $(\unit[1.78] x \unit[1.78]){\mu m}^2$, centered around $(\unit[0,0]){\mu m}$ in Figs. (a-c), comparable with the area of the $2~\mu m$ FWHM excitation spot used in the experiment. The letters (A), (B) and (C) in the graph refer to Figs. (a-c) }
\label{fig:C_stokes}
\vspace{0.2cm}
\end{figure}
In the case of circularly polarized excitation shown in Fig. \ref{fig:C_stokes}, polariton condensation results in highly imbalanced population [Fig.\,\ref{fig:5} (b)] and the small ellipticity induced by the tightly focused spot will not play a relevant role as in the case of linearly polarized pump. As a consequence, the imbalance between the two polariton populations is set by the pump and preserved throughout the entire process, so that the polarization do not rotate at the pump spot. Therefore, no oscillation in the density [Fig.\,\ref{fig:5} (b)] and in the polarization at the pump spot [Fig.\ref{fig:C_stokes} (d)] appear and the orientation of the four-leaf clover pattern typical of the OSHE, remain fixed in time (i.e the spin texture does not rotate). %At early time, the pseudospin vector at the pump position is collinear to the $z$-direction and to the $\Omega_z$ field and, consequently, no precession occurs at the pump spot. Therefore, the dynamics is mainly dominated by $\Omega_{LT}$ and the collective rotation of the spin in the plane of the microcavity manifests itself as tails in the $H$ component (red in Fig.\,\ref{fig:C_stokes}), connecting the inner and external domains. 
As polaritons propagate radially outward in the plane of the microcavity, their spin precesses around $\Omega_{LT}$, giving rise to characteristic shape domains already observed in Ref.\,\cite{kammann_nonlinear_2012}. %At higher excitation power ($\unit[20]{m W}$), the effect becomes more visible due to greater density of polaritons injected into the cavity (see supplementary video S4b) {\redc \bf (It seems to me that at higher pumping power the spin lobes disconnect faster. This is good since I see the same if the pumping power is increase for circular pumping)}. On the other hand, in the case of circularly polarized excitation and circular detection (Fig.\ref{fig.th.pulse} (a)), the total symmetric circular textures prevents the possibility to observe a rotational effect. 

In the simulations shown in Fig. \ref{fig:Th_C_stokes}, we use a nearly circular pulse, where the ellipticity is set to $\sigma_+ = 1, \  \sigma_- = 0.1$. The other parameters were set to: $\alpha = 2.4$ $\mu$eV $\mu$m$^2$, $g_R = 1.5 \alpha$, $G = 4 \alpha$, $r_c = 0.01$ $\mu$m$^2$ ps$^{-1}$, $\Delta_{LT}/k_{LT}^2 = 11.9$ $\mu$eV $\mu$m$^2$, $\tau_p = 9$ ps, $\tau_x = 10$ ps. %However, here the ellipticity introduced by the high NA objective does not play a relevant role because a circularly polarized excitation results in a highly unbalanced condensate.  %The linear polarization shows spin tails connecting the inner polarization domain to the outer one like in the experimental (Fig. \ref{fig:C_stokes}).%(\emph{Comment: This needs a bit more work.}).
\begin{figure}[!hbtp]
\includegraphics[scale=0.6]{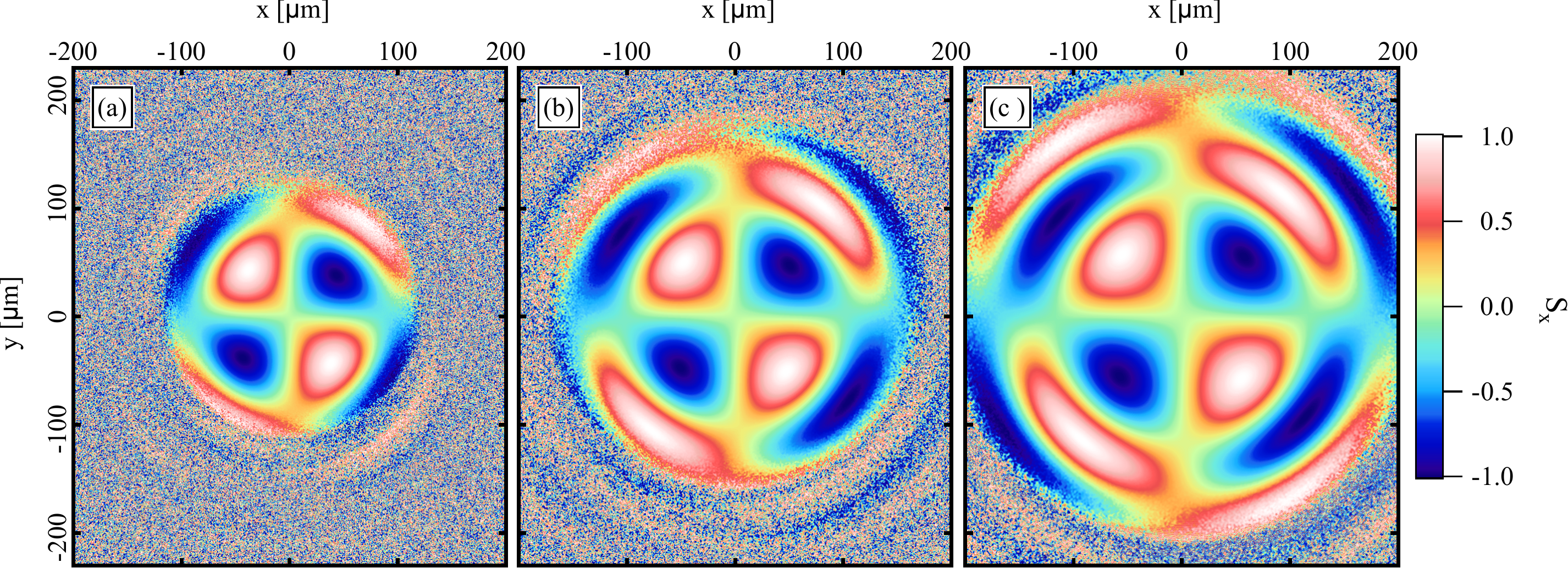}
\caption{Theoretical linear Stokes parameters at (a) $\unit[40]{ps}$, (b) $\unit[55]{ps}$ and (c) $\unit[70]{ps}$ showing the formation and evolution of the linear Stokes spin texture in the plane of the microcavity. The polarization of the excitation beam is set to $\sigma_+ = 1, \  \sigma_- = 0.1$.} %The parameters used are: $\alpha = 2.4$ $\mu$eV $\mu$m$^2$, $g_R = 1.5 \alpha$, $G = 4 \alpha$, $r_c = 0.01$ $\mu$m$^2$ ps$^{-1}$, $\Delta_{LT}/k_{LT}^2 = 11.9$ $\mu$eV $\mu$m$^2$, $\tau_p = 9$ ps, $\tau_x = 10$ ps.}
\label{fig:Th_C_stokes}
\end{figure}
\end{widetext}

\newpage
\bibliography{Bibliography}
\end{document}